\documentclass[twoside]{article}
\usepackage{fleqn,espcrc2}

\usepackage{epsf}

\usepackage{latexsym}

\newcommand{\AmS}{{\protect\the\textfont2
  A\kern-.1667em\lower.5ex\hbox{M}\kern-.125emS}}

\title{$K\to\pi\pi$ Decays with Domain Wall Fermions: Lattice Matrix Elements}

\author{T. Blum\address{RIKEN-BNL Research Center,
Brookhaven National Laboratory, Upton, NY 11973-5000, USA}\\
{RBC Collaboration}
\thanks{This work was done in collaboration with
N.~Christ, C.~Cristian, C.~Dawson, G.~Fleming,
X.~Liao, G.~Liu, S.~Ohta, A.~Soni, P.~Vranas, R.~Mawhinney,
M.~Wingate, L.~Wu, and
Y.~Zhestkov.  We thank RIKEN, Brookhaven National Laboratory and the
U.S.\ Department of Energy for providing the facilities essential for
this work.}}

\newcommand{\bea}{\begin{eqnarray}}
\newcommand{\eea}{\end{eqnarray}}
\newcommand{\EL}{\nonumber\\}
       
\begin{document}

\begin{abstract}
We present a lattice calculation of the $K\to\pi$ and $K\to 0$ matrix
elements of the $\Delta S=1$ effective weak Hamiltonian which can be used to
determine
$\epsilon^\prime/\epsilon$ and the $\Delta I=1/2$
rule for $K$ decays in the Standard Model. The matrix elements
for $K\to\pi\pi$ decays are related to $K\to\pi$ and $K\to 0$ using
lowest order chiral perturbation theory. We also
present results for the kaon $B$ parameter, $B_K$. Our quenched domain
wall fermion
simulation was done at $\beta=6.0$ ($a^{-1}\approx 2$ GeV), 
lattice size $16^3\times 32\times 16$,
and domain wall height $M_5=1.8$. 
\vspace{1pc}
\end{abstract}
\maketitle

\section{Introduction}
Recent measurements of direct $CP$ violation ($\epsilon^\prime/\epsilon\neq 0)$
in $K\to\pi\pi$ decays at FNAL and CERN allow an important
test of the Standard Model (in particular, the $CKM$ mixing paradigm).
The effective weak Hamiltonian governing strangeness changing
$K$ decays has been computed 
to next-to-leading order in QCD and QED by the Munich and Rome groups; 
the remaining 
piece of the puzzle is the hadronic matrix elements of the
operators of this effective weak Hamiltonian.

The recent advance of domain wall (and overlap) fermions which maintain
chiral symmetry to a high degree of accuracy\cite{DWF} allows for a
new attempt at this old problem. Chiral symmetry of domain wall
fermions provides a significant advantage when computing 
light quark QCD observables since the lattice artifacts that
arise when this symmetry is explicitly broken are greatly reduced.
Mixing and renormalization of operators, which is already complicated
in the continuum, is readily handled with domain wall fermions\cite{Chris}.
However, calculations using improved Wilson fermions 
were also reported at this meeting\cite{Guido}, and it is still unclear
which method will prove most advantageous.

In this study, we present preliminary results for the
$K\to\pi$ and $K\to 0$ matrix elements of these operators,
which when combined with lowest order chiral perturbation theory,
yield the desired $K\to\pi\pi$ matrix elements\cite{Bernard}.
The contribution of R. Mawhinney in these proceedings takes up
this point\cite{Bob}. Here we are concerned with the simpler 
$K\to\pi$ and $K\to 0$ lattice
matrix elements. Also, the CP-PACS collaboration has presented a very
similar calculation at this meeting\cite{Noaki}.

\section{Theoretical Framework}

The $\Delta S=1$ effective weak Hamiltonian is
generated from the fundamental Standard Model Lagrangian 
by integrating out the top quark and W boson. The resulting
effective Hamiltonian is then
evolved to a low scale ($\mu\ll M_W$) appropriate for lattice
calculations
using the renormalization group equations. The effective Hamiltonian
above the charm threshold is:
\bea
H^{\Delta S=1}&=&V_{ud}V^*_{us}\frac{G_F}{\sqrt{2}}
\left [
\left(1+\frac{V_{td}V^*_{ts}}{V_{ud}V^*_{us}}\right)\right.\EL
&&(C_1(\mu)({Q_1(\mu)}-{Q_{1c}(\mu)})\EL &+&
C_2(\mu)({Q_2(\mu)}-{Q_{2c}(\mu)}))\EL
&-&  \left.
\frac{V_{td}V^*_{ts}}{V_{ud}V^*_{us}} \vec C(\mu) \cdot 
{\vec Q(\mu)}\right].
\eea
where $\mu$ is the renormalization scale,
$\vec Q$ are 
a basis of local four-quark operators 
which are closed under 
renormalization,
$C_i(\mu)$ the corresponding 
Wilson coefficients, 
and $V_{qq^\prime}$ the
Cabibbo-Kobayashi-Maskawa mixing matrix elements
which are fundamental parameters of the Standard Model.

The effective operators
renormalized at the scale $\mu$ are 
\bea
\label{basis}
Q_1&=& 
	{\bar s_\alpha} 
	\gamma_\nu P_L 
	{d_\alpha}
	{\bar u_\beta} 
	\gamma_\nu P_L
	{ u_\beta}\EL
Q_2&=&
        {\bar s_\alpha}
        \gamma_\nu P_L
        {d_\beta}
        {\bar u_\beta} 
        \gamma_\nu P_L
        { u_\alpha}\EL
Q_{3,5}&=& 
        {\bar s_\alpha}
        \gamma_\nu P_L
        {d_\alpha}
        {\sum_{u,d,s,c\dots}\bar q_\beta} 
        \gamma_\mu \gamma_\nu P_{(L,R)}
        { q_\beta}\EL
Q_{4,6}&=&
        {\bar s_\alpha}
        \gamma_\nu P_L
        {d_\beta}
        {\sum_{u,d,s,c\dots}\bar q_\beta}
        \gamma_\mu \gamma_\nu P_{(L,R)}
        { q_\alpha}\EL
Q_{7,9}&=& \frac{3}{2}
        {\bar s_\alpha}
        \gamma_\nu P_L
        {d_\alpha}
        {\sum_{u,d,s,c\dots}e_q\bar q_\beta}
        \gamma_\mu \gamma_\nu P_{(R,L)}
        { q_\beta}\EL
Q_{8,10}&=& \frac{3}{2}
        {\bar s_\alpha}
        \gamma_\nu P_L
        {d_\beta}
        {\sum_{u,d,s,c\dots}e_q\bar q_\beta}
        \gamma_\mu \gamma_\nu P_{(R,L)}
        { q_\alpha}\EL
Q_{1c}&=&
        {\bar s_\alpha}
        \gamma_\nu P_L
        {d_\alpha}
        {\bar c_\beta}
        \gamma_\nu P_L
        { c_\beta}\EL
Q_{2c}&=&
        {\bar s_\alpha}
        \gamma_\nu P_L
        {d_\beta}
        {\bar c_\beta} 
        \gamma_\nu P_L
        { c_\alpha},
\eea
with color indices
$\alpha$ and $\beta$,
$P_{(L,R)}=1\mp\gamma_5$,
the sums are taken over active quark flavors at the scale $\mu$,
and summation over $\nu$ is implied.
$Q_{1,2,1c,2c}$ are often referred to as
current-current operators, $Q_{3-6}$ QCD penguin operators,
and $Q_{7-10}$ electroweak penguin operators.
It is useful to split the above operators according to their isospin,
$Q_i\equiv Q_i^{(1/2)} + Q_i^{(3/2)}$.

As mentioned earlier, lowest order chiral perturbation theory relates
$K\to\pi\pi$ matrix elements to a linear combination of
$K\to\pi$ and $K\to 0$. For all $Q_i$ except the
electroweak penguins $Q_{7,8}$ and all particles at rest, 
we have\cite{Bernard}
\bea
{\langle \pi^+\pi^-|Q|K^0\rangle} &=& 
 \frac{4i(m_K^2-m_{\pi}^2)\alpha_1}{f^3}\EL
{\langle \pi^+| Q|K^+\rangle} &=& 
 \frac{4m_M^2(\alpha_1-\alpha_2)}{f^4}\EL
{\langle0| Q|K^0\rangle} &=& 
 \frac{4i(m_K^2-m_{\pi}^2)\alpha_2}{f},
\eea
where $m_M$ is the meson mass for unphysical pseudoscalar states
with $m_s=m_d$. 
For $Q_i$ which transform in a (27,1) chiral multiplet $\alpha_2=0$.
Note that each matrix element
vanishes linearly with
the meson mass squared. This is an important prediction of
chiral perturbation theory, and therefore QCD, and provides
a solid test of the chiral symmetry properties of domain wall
fermions. The strength of the above approach is that it allows less
computationally demanding 
$K\to\pi$ and $K\to 0$ matrix elements to be calculated on the
lattice. A significant drawback to this approach is that it
manifestly does not contain information on the final state interactions
of the pions (for calculation of $\epsilon^\prime$ the final state
s-wave scattering phases from experiment can be put in by hand, however).

In the case of the electroweak
penguins $Q_{7,8}$, the contribution in lowest order chiral perturbation
theory to the $K\to\pi$ matrix elements 
is constant in the chiral limit. The $K\to 0$ matrix element, however,
still vanishes.

Since on the lattice $\alpha_2$ is quadratically divergent,
the process of combining the second and third
lines in Eq. 4 to obtain an expression for
$K \to \pi\pi$ requires a delicate cancellation
of this quadratic divergence.  See R. Mawhinney's
contribution for details.

Finally, lattice counterparts of the operators in Eq.~\ref{basis} 
must be matched to the continuum and renormalized since
they are logarithmically divergent after power divergences have been 
subtracted. In addition, operators in the same symmetry multiplets
mix through renormalization group running from $M_W$ down to 
the low scale $\mu$.
This poses a serious challenge for lattice calculations.
In our calculation this problem is handled remarkably well
with the nonperturbative renormalization method of the Rome-Southampton
group
which was explained in the talk by C. Dawson\cite{Chris}.

\section{Simulation details}

We have calculated matrix elements on 200 quenched gauge configurations
at $\beta=6.0$, with lattice four volume $16^3\times 32$, domain wall
fermion extra dimension size $L_s=16$, and domain
wall height $M_5=1.8$. We have calculated with light
quark masses $m_f=0.01-0.05$, and charm quark masses $m_c=0.1-0.4$. The
physical kaon state made from degenerate quarks corresponds to 
$m_f\approx 0.02$, and $m_c\approx 0.5$ for the physical charm quark.

We extract matrix elements from three-point correlation functions.
The external pseudoscalar states are interpolated from wall sources
near the time direction boundaries, $t=5$ and $27$,
and the operator is inserted
between them. When the operator is far from either boundary, the
desired lowest mass states dominate the correlation function.
The forward and backward (in time) quark propagators used to
interpolate the $K$ and $\pi$ states
are linear combinations of propagators computed with periodic and
anti-periodic boundary conditions which amounts to doubling the
gauge field configuration in the time direction.
The closed fermion loops
necessary for operators that have self contractions are computed
from a complex Gaussian random source spread over time slices 14-17.
All results are given as averages over these four time slices. 

We have performed several important checks of our computer code. Most
importantly, a completely independent check code was written to
compare with our two production versions (the check code and one
of the production codes run on the QCDSP supercomputer and the
other production code, based on the MILC code,
runs on the NERSC T3E). Output from each code generated on the
same configuration agreed up to machine precision.
  
As a final useful check, 
the left-left operators in Eq.~\ref{basis} go
into themselves under a Fierz transformation. Thus color-mixed
contractions can be compared to corresponding color-diagonal ones. We find
perfect agreement in all cases.

The following results were obtained on the RIKEN BNL and Columbia University
QCDSP supercomputers.
\section{Results}
 
\begin{figure}[hbt]
    \vskip -.2in
    \vbox{ \hskip -.35in\epsfxsize=3.0in \epsfbox[0 0 4096 4096]{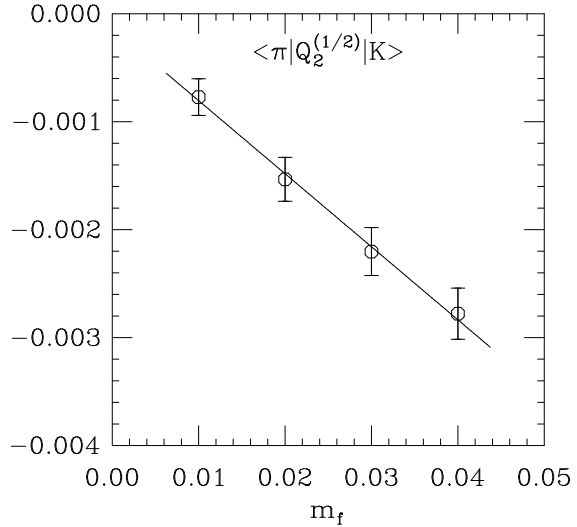} }
    \vskip -.4in
    \caption{The $K\to\pi$ matrix element of the bare operator $Q^{(1/2)}_2$.}
    \label{lato2}
\end{figure}

\begin{figure}[hbt]
    \vskip -.2in
    \vbox{ \hskip -.25in\epsfxsize=3.0in \epsfbox[0 0 4096 4096]{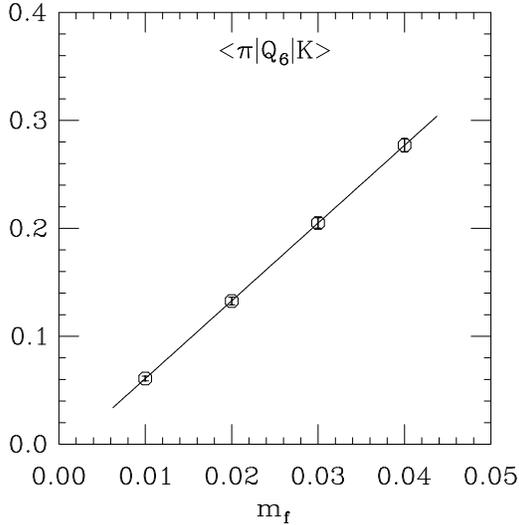} }
    \vskip -.4in
    \caption{The $K\to\pi$ matrix element of the bare operator $Q_6$ (charm
             contribution not included).}
    \label{lato6}
\end{figure}

\begin{figure}[hbt]
    \vskip -.2in
    \vbox{ \hskip -.25in\epsfxsize=3.0in \epsfbox[0 0 4096 4096]{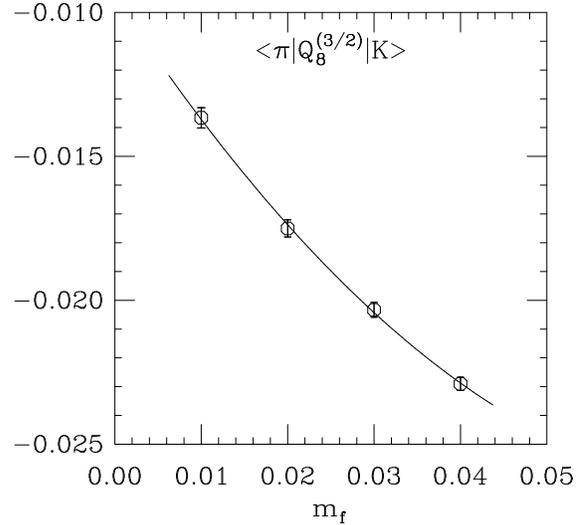} }
    \vskip -.4in
    \caption{The $K\to\pi$ matrix element of the bare operator $Q^{(3/2)}_8$.}
    \label{lato18}
\end{figure}

Fig.~\ref{lato2} shows $\langle\pi|Q^{(1/2)}_2|K\rangle$ as a function of
quark mass, $m_f=m_s=m_d$. An uncorrelated 
linear extrapolation yields a zero intercept, within
statistical errors, which is in
agreement with chiral perturbation theory. For strictly low energy QCD 
observables, we expect quantities to vanish at $m_f=-m_{res}\cite{DWF}$. Since
the $\Delta I=1/2$ operators have contributions from physics scales near
the (high energy) lattice cut-off, this is no longer true. Thus, for this
matrix element, the statistical errors are not small enough to resolve 
these systematic effects. However, in R. Mawhinney's contribution, we
see that such
effects are visible in the {\it subtracted} operator\cite{Bob}. 
Presumably this is
due to the strong statistical 
correlations between the $K\to\pi$ matrix element and
the subtraction term.
In Fig.~\ref{lato6} we show a similar plot for $Q_6$. Here, an uncorrelated
linear extrapolation has a non-zero intercept of roughly three standard 
deviations. Note that it vanishes for $m_f>0$. Thus explicit chiral symmetry
breaking effects are visible, though small.
In Fig.~\ref{lato18} we show $\langle \pi|Q^{(3/2)}_8|K\rangle$ which exhibits
noticeable nonlinearity and does not vanish in the chiral limit, as expected.

\begin{figure}[h!]
    \vskip -.2in
    \vbox{ \hskip -.2in\epsfxsize=3.0in \epsfbox[0 0 4096 4096]{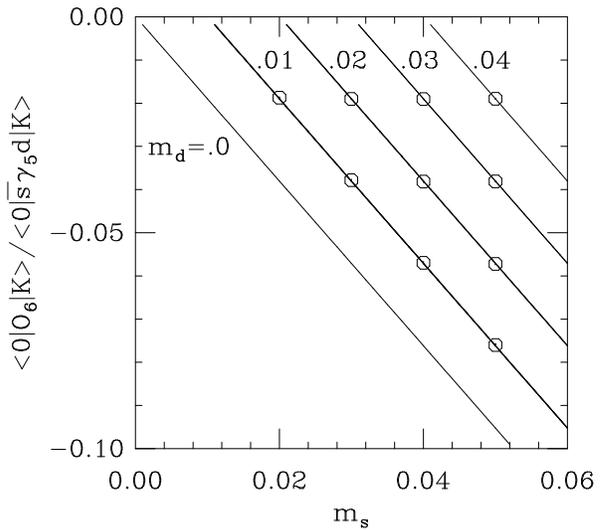} }
    \vskip -.4in
    \caption{The $K\to 0$ matrix element of the bare operator $Q_6$ (charm
             contribution not included).}
    \label{vaco6}
\end{figure}

Finally, we show an example of a $K\to 0$ matrix element in
Fig.~\ref{vaco6}. Note that for the $K\to 0$ matrix elements
we use $m_s\neq m_d$ and fit to the form
$\langle 0|Q_i|K\rangle/\langle 0|\bar{s}\gamma_5 d|K\rangle=const+
(m_s-m_d)\eta_i$ where chiral perturbation theory predicts $const =0$.
This ratio is useful since it is exactly the coefficient of the
subtraction operator used to remove the quadratic divergence in 
$\langle \pi|Q_i|K\rangle$.
Since the quark masses enter as a difference, we expect explicit
chiral symmetry breaking effects which do not depend on the quark mass
to cancel. 
From the fit depicted in Fig.~\ref{vaco6} we find that the
constant term is zero within errors. Note that this ratio is
extremely well resolved, and quite linear.

We take this opportunity to quote our value for 
the kaon $B$ parameter,
$B^{(\overline{MS})}_K(2\, {\rm GeV})=0.538(8)$, 
with $Z_{LL}/Z^2_A=0.928(6)$ computed
nonperturbatively in the RI scheme and matched to the
$\overline{MS}$-$NDR$ scheme\cite{Crisafulli}. 
The errors are statistical only,
and the error on $B_K$ is obtained by adding the errors on the
matrix element and the renormalization factor in quadrature.
Our value is lower
than the one quoted by Taniguchi at this meeting\cite{B_K}, 
probably due to the fact that our renormalization
constant is lower than the perturbative one used in that study.
Note our result is for $\beta=6.0$ with a single lattice size.

\end{document}